\documentclass[10pt,aps,prl,showpacs,superscriptaddress,nofootinbib,twocolumn,preprintnumbers,floatfix]{revtex4-1}
\usepackage{graphicx,amsmath,amssymb,soul,enumitem,bbm,dcolumn,float}
\usepackage{booktabs,multirow,array,slashed,dsfont,subcaption,mwe}
\usepackage[dvipsnames]{xcolor}
\usepackage{nicefrac}
\usepackage[utf8]{inputenc}
\allowdisplaybreaks
\usepackage{hyperref}
\usepackage{tabularx}
\hypersetup{colorlinks=true,urlcolor=PineGreen,linkcolor=PineGreen,citecolor=PineGreen}

\makeatletter
\newcommand\nocaption{%
    \renewcommand\p@subfigure{}
    \renewcommand\thesubfigure{\thefigure\alph{subfigure}}
}
\makeatother

\begin{document}

\title{
S- and p-wave structure of $S=-1$ meson-baryon scattering in the resonance region
}

\author{D.\ Sadasivan}
\email{dansadasivan@gwu.edu}
\affiliation{Institute for Nuclear Studies and Department of Physics, The George Washington University, Washington, DC 20052, USA}
\author{M.\ Mai}
\email{maximmai@gwu.edu}
\affiliation{Institute for Nuclear Studies and Department of Physics, The George Washington University, Washington, DC 20052, USA}
\author{M.\ D\"oring}
\email{doring@gwu.edu}
\affiliation{Institute for Nuclear Studies and Department of Physics, The George Washington University, Washington, DC 20052, USA}
\affiliation{Thomas Jefferson National Accelerator Facility, Newport News, VA 23606, USA}

\preprint{JLAB-THY-18-2699}

\begin{abstract}
We perform a simultaneous analysis of s- and p-waves of the $S=-1$ meson-baryon scattering amplitude using all low-energy experimental data. For the first time, differential cross section data are included for chiral unitary coupled-channel models. From this model s- and p-wave amplitudes are extracted and we observe both well-known $I(J^P)=0(1/2^-)$ s-wave states as well as a new $I(J^P)=1(1/2^+)$ state absent in quark models and lattice QCD results. Multiple statistical and phenomenological tests suggest that, while the data clearly require an $I=1$ p-wave resonance, the new state just accounts for the absence of the decuplet $\Sigma(1385)3/2^+$ in the model. 
\end{abstract}

\pacs{
11.30.Rd, 
11.80.Gw, 
14.20.Jn  
}

\maketitle

\paragraph{\textbf{Introduction}}

Strangeness plays an important role in various facets of strongly interacting matter. For example, the attractive interaction between antikaons and nucleons leads to the famous $\Lambda(1405)$-resonance as predicted in Ref.~\cite{Dalitz:1959dn}. Nowaday this low-energy, $S=-1$ region is accessed by chiral unitary coupled-channel models~\cite{Mai:2014xna, Mai:2012dt, Cieply:2016jby, Miyahara:2018onh, Liu:2016wxq, Ramos:2016odk, Kamiya:2016jqc,  Feijoo:2015yja, Molina:2015uqp, Guo:2012vv, 
Doring:2011ip, Ikeda:2011pi, Cieply:2011nq,Cieply:2015pwa, Jido:2002zk,
Doring:2010rd, Oller:2006jw, Oset:2001cn, Oller:2000fj,  Oset:1997it}, amplitude analyses~\cite{Fernandez-Ramirez:2015tfa, Fernandez-Ramirez:2015fbq, Kamano:2015hxa, Kamano:2014zba, Zhang:2013sva, Zhang:2013cua}, lattice QCD~\cite{Hall:2016kou, Hall:2014uca, Engel:2013ig, Edwards:2012fx, Engel:2010my}, and quark models~\cite{Xie:2014zga, Helminen:2000jb, Jaffe:2003sg}, see, e.g., the reviews in Refs.~\cite{Guo:2017jvc, Tanabashi:2018oca}. A similar mechanism can be responsible for the generation of $K^-pp$ bound states~\cite{Ajimura:2018iyx,Sada:2016nkb} as predicted in Ref.~\cite{Sekihara:2016vyd}. Furthermore, the equation of state of neutron stars is sensitive to the antikaon condensate~\cite{Kaplan:1986yq, Pal:2000pb} and thus to the propagation of antikaons in nuclear medium. In the era of high-precision measurements of neutron star properties with LIGO~\cite{Abadie:2011xta}, this ultimately can lead to new interconnection between QCD and astrophysical observations. 

At the core of all theoretical studies lies the antikaon-nucleon scattering amplitude, which has to address the non-perturbative regime of QCD valid over a large energy range including the resonance region. Such a "twice non-perturbative" amplitude can neither rely on perturbative QCD nor its low-energy effective field theory. Thus, some model dependence has to be introduced, with the corresponding parameters being fitted to experimental data. In this work we use the model derived in a series of works~\cite{Mai:2014xna,Mai:2013cka,Mai:2014xna,Bruns:2010sv} which to our knowledge is the only approach, which has the correct low-energy behavior, fulfills two-body unitarity and describes s- and p-waves simultaneously without introducing explicit states. This property is utilized to fit all available scattering and threshold data including (for the first time) differential cross sections, simultaneously, up to energies well below the d-wave $\Lambda(1520)3/2^-$ resonance.
\paragraph{\textbf{Model}}

\renewcommand\arraystretch{1.3}
\begin{table*}[thb!]
\begin{tabular}{|c|c|}
\hline
$\chi^2_{\rm dof}$ &$1.10~(62/30/8\%)$\\
~~~~~~~$|\Sigma|~[{\rm GeV^{-14}}]$~~~~~~~~	&$0.106$\\
\hline
\multicolumn{2}{c}{~}\\[-6pt]
\hline
$a_{1-6}$&$b_{1-11}~[{\rm GeV^{-1}}]$\\
\hline
$+0.44^{+0.28}_{-0.14}$&$-0.34^{+0.11}_{-0.04}$\\
$+2.04^{+1.37}_{-0.50}$&$+0.29^{+0.21}_{-0.18}$\\
$+0.14^{+0.15}_{-0.20}$&$-0.62^{+0.04}_{-0.06}$\\
$-0.95^{+0.30}_{-0.31}$&$-0.02^{+0.16}_{-0.14}$\\
$-0.75^{+0.16}_{-0.45}$&$+0.24^{+0.06}_{-0.11}$\\
$-1.99^{+0.30}_{-3.59}$&$-0.82^{+0.31}_{-0.09}$\\
		&$-1.32^{+0.44}_{-0.04}$\\
\cline{1-1}
$b_{0,D,F}~[{\rm GeV^{-1}}]$	&$-0.01^{+0.02}_{-0.10}$\\
\cline{1-1}
$-0.50^{+0.01}_{-0.01}$&$+0.04^{+0.22}_{-0.14}$\\
$+0.08^{+0.01}_{-0.01}$&$+0.15^{+0.05}_{-0.06}$\\
$-0.22^{+0.01}_{-0.01} $&$+0.40^{+0.13}_{-0.05}$\\
		\hline
\end{tabular}
~~~~
\begin{tabular}{|l|l|c|c|c|}
\hline 
\multicolumn{2}{|c|}{$I(J^P)$}&$0(1/2^-)$&$0(1/2^-)$&$1(1/2^+)$\\
\hline
\multicolumn{2}{|c|}{Pole position [MeV] }&$1430-15i$&$1360-43i$&$1360-11i$\\[5pt]
\multicolumn{2}{|c|}{Covariance [MeV$^2$]}&$\begin{pmatrix}+20.0&+3.7\\&+14.9\end{pmatrix}$ 
&$\begin{pmatrix}+169&~-166\\~&~+202\end{pmatrix}$ 
&$\begin{pmatrix}+354&~-91\\~&~+31\end{pmatrix}$\\[10pt]
\hline
\multirow{10}{*}{\rotatebox{90}{Couplings [GeV]}}
&~~~~$g_{K^-p}$	&$0.34^{+0.10}_{-0.05}$&$0.35^{+0.06}_{-0.03}$&$0.10^{+0.05}_{-0.03}$\\
&~~~~$g_{\bar K^0 n}$	&$0.32^{+0.10}_{-0.05}$&$0.35^{+0.06}_{-0.04}$&$0.10^{+0.04}_{-0.03}$\\
&~~~~$g_{\pi^0\Lambda}$	&$-$		       &$-$		      &$0.19^{+0.03}_{-0.03}$\\
&~~~~$g_{\pi^0\Sigma^0}$	&$0.19^{+0.03}_{-0.02}$&$0.39^{+0.04}_{-0.03}$&$-$\\
&~~~~$g_{\pi^+\Sigma^-}$	&$0.19^{+0.03}_{-0.02}$&$0.40^{+0.04}_{-0.03}$&$0.11^{+0.06}_{-0.07}$\\
&~~~~$g_{\pi^-\Sigma^+}$	&$0.19^{+0.03}_{-0.02}$&$0.38^{+0.04}_{-0.03}$&$0.09^{+0.06}_{-0.06}$\\
&~~~~$g_{\eta\Lambda}$	&$0.22^{+0.23}_{-0.09}$&$0.30^{+0.10}_{-0.07}$&$-$\\
&~~~~$g_{\eta \Sigma^0}$	&$-$		       &$-$	              &$0.13^{+0.25}_{-0.07}$\\
&~~~~$g_{K^+\Xi^-}$	&$0.05^{+0.03}_{-0.03}$&$0.02^{+0.02}_{-0.01}$&$0.08^{+0.07}_{-0.06}$\\
&~~~~$g_{K^0\Xi^0}$&$0.06^{+0.03}_{-0.03}$&$0.02^{+0.02}_{-0.01}$&$0.08^{+0.07}_{-0.06}$ \\
\hline
\end{tabular}
\caption{
\underline{Left}: Parameters of the best fit. The values in parentheses behind the $\chi^2{\rm dof}$ show the contributions from data sets (a,b,c), respectively. The generalized variance is given by the determinant of the covariance matrix, $|\Sigma|$, and is related to the volume of the error ellipse as described in the main text. Error bars on the model parameters are determined in a re-sampling procedure. \underline{Right}: Prediction of resonance parameters found in a channel of given isospin ($I$) and total angular momentum ($J=L\!\pm\!\nicefrac{1}{2}$) on the second Riemann sheet, including the absolute values of the couplings to meson-baryon channels.
}
\label{tab:fit-results}
\end{table*}

The entire database for meson-baryon scattering in the strangeness $S=-1$ sector comes from experiments with $K^-p$ in the initial state. In general, 10 combinations of ground state octet mesons and baryons have the same quantum numbers, i.e. $\mathcal{S}:=\{K^-p$, $\bar K^0 n$, $\pi^0\Lambda$, $\pi^0\Sigma^0$, $\pi^+\Sigma^-$, $\pi^-\Sigma^+$, $\eta\Lambda$, $\eta \Sigma^0$, $K^+\Xi^-$, $K^0\Xi^0\}$ meaning that the scattering amplitude $T$ must describe the dynamics of all coupled-channels simultaneously. Two-body unitarity constrains the form of such a scattering amplitude, incorporated exactly via the Bethe-Salpeter equation in $d$ Minkowski dimensions
\begin{align}
\label{eqn:BSE}
T(&\slashed{q}_2, \slashed{q}_1; p)=
V(\slashed{q}_2, \slashed{q}_1;p)\\
&+
i\int\frac{d^d \ell}{(2\pi)^d}
\frac{V(\slashed{q}_2, \slashed{\ell}; p)}{\ell^2-M^2+i\epsilon} 
\frac{1}{\slashed{p}-\slashed{\ell}-m+i\epsilon}
T(\slashed{\ell}, \slashed{q}_1; p)\,,\nonumber
\end{align}
where $p$ is the total and $q_{1/2}$ are the in-/outgoing meson four-momenta, while $m/M$ denotes the mass of the baryon/meson in each channel, respectively. Different channels are related by SU$(3)_{\rm f}$ symmetry and its breaking in the interaction kernel $V$ and can be further restricted by demanding the correct energy-dependence around the corresponding thresholds. This requirement is addressed by taking the ChPT~\cite{Krause:1990xc,Frink:2004ic} local potential up to the next-to-leading chiral order
\begin{align}\label{eqn:potential}
&V(\slashed{q}_2, \slashed{q}_1; p)=A_{WT}(\slashed{q_1}+\slashed{q_2})
+A_{1-4}(q_1\cdot q_2)\\
&+A_{5-7}[\slashed{q_1},\slashed{q_2}]
+A_{0DF} +A_{8-11}\Big(\slashed{q_2}(q_1\cdot p)+\slashed{q_1}(q_2\cdot p)\Big)\,,\nonumber
\end{align} 
where $A$ are matrices over the channel space that depend on the meson decay constants (fixed throughout this work together with meson and baryon masses to the physical values~\cite{Tanabashi:2018oca}) and low-energy constants (LECs) $\{b_0,b_D,b_F,b_1,...,b_{11}\}$. The latter are not known from ChPT and must be fitted to the experimental data.

The present model corresponds to an infinite set of ChPT Feynman diagrams. Since this set does not contain all possible diagrams, the regularization scale of the loop integration in Eq.~\eqref{eqn:BSE} does not cancel out. To express it differently, this scale  parametrizes the missing Feynman topologies to some extent and is dealt with as free parameters of the model. Neglecting isospin breaking, there are six such parameters, called $\{a_i|i=1,..,6\}$ in the following for brevity. Previous analyses~\cite{Mai:2013cka,Bruns:2010sv,Mai:2012dt} have shown that off-shell terms explicitly contained in Eq.~\eqref{eqn:BSE} have only minor effect and are dropped here as well. With this approximation the scattering amplitude is solved analytically and is projected to physical observables~\cite{Hohler:1984ux}. In particular, for channel indices $i$ and $j$ the partial-wave amplitudes read
\begin{align}
16&\pi W f_{L\pm}^{ij}=\\
+&\sqrt{E_i+m_i}\sqrt{E_j+m_j}\left(A_L^{ij}+\left(W-\frac{m_i+m_j}{2}\right)B_L^{ij}\right)\nonumber\\
-&\sqrt{E_i-m_i}\sqrt{E_j-m_j}\left(A_{L\pm1}^{ij}-\left(W+\frac{m_i+m_j}{2}\right)B_{L\pm1}^{ij}\right)\,.\nonumber
\end{align}
Here $E=\sqrt{m^2+q_{cms}^2}$ for $q_{cms}$ being the modulus of the three-momentum in the center of mass system, while $A_L^{ij}$ and $B_{L}^{ij}$ denote the partial wave projected (to angular momentum $L$) invariant amplitudes. On shell the latter are related to the scattering amplitude~\eqref{eqn:BSE} as $T^{ij}_{\rm ON}=A^{ij}+(\slashed{q}+\slashed{q'})B^{ij}$, see for more details and further definitions Sec.~2 of Ref~\cite{Hohler:1984ux}. To our knowledge this model is the only existing unitary coupled-channel model which contains explicit s- and p-wave interactions, derived from the low-energy behavior of QCD Green's functions.

\paragraph{\textbf{Data and fits}}

\begin{figure*}[t]
\includegraphics[width=\linewidth, trim = 0 0.2cm 0 0]{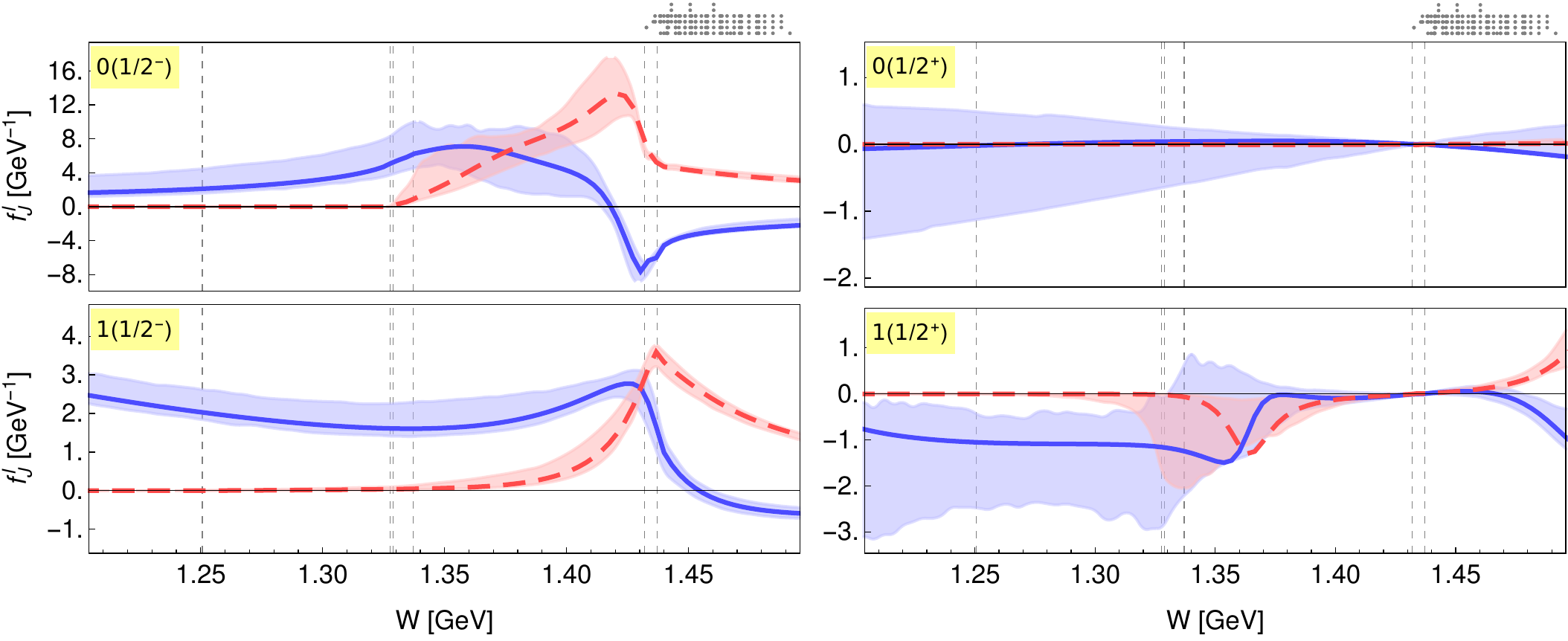}
\caption{
Real (solid, blue) and imaginary (red, dashed) parts with $1\sigma$ uncertainty bands of $\bar KN$ partial wave amplitudes with $I(J^P)$. The vertical dashed lines show the positions of the $\pi\Lambda$, $\pi^0\Sigma^0$, $\pi^+\Sigma^-$, $\pi^-\Sigma^+$, $K^-p$, $\bar K^0n$ thresholds, and the dots over the plots represent the available data in the respective channel from top to bottom.
}
\label{fig:amplitudes}
\end{figure*}

The available data consist of: (a) Total cross sections~\cite{Ciborowski:1982et,Humphrey:1962zz,Sakitt:1965kh,Watson:1963zz} for $\{K^-p\to \mathcal{S}_i\,|i=1,..,6\}$ in the lab-momentum range of $P_{lab}<300$~MeV; (b) Differential cross sections~\cite{Mast:1975pv} in the same momentum window for the $K^-p\to K^-p$ and $K^-p\to\bar K^0n$ reactions; (c) 
Threshold decay ratios~\cite{Tovee:1971ga,Nowak:1978au} $\gamma=2.38\pm0.04$, $R_n=0.189\pm0.015$, $R_c=0.664\pm0.011$, and energy shift $\Delta E=283\pm42$~eV and width $\Gamma=541\pm110$~eV of kaonic hydrogen measured in the SIDDHARTA~\cite{Bazzi:2011zj} experiment and related~\cite{Meissner:2004jr} to the $K^-p$ complex-valued scattering length $a_{K^-p}$.
Further experimental data, related indirectly to the $\bar KN$ scattering amplitude, require additional model assumptions and are not used in fits. \textit{A-posteriori} checks to these data will be discussed below.  

We employ a multi-step fitting routine to explore the parameter space as extensively as possible. \underline{First}, we set all parameters $b$ to zero, effectively reducing the chiral order of the driving term $V$ of Eq.~\eqref{eqn:BSE}. The remaining parameters are fitted to the data starting from a large set $\mathcal{O}(10^3)$ of randomized starting values of natural size, i.e. $|a_i|\lesssim5$. \underline{Second}, using the best fits of the previous step as starting values we fit all parameters of the model simultaneously. The starting values of the so-called dynamical LECs $b_1-b_{11}$ are chosen randomly and fitted within natural limits $|b_{1-11}|<10~{\rm GeV}^{-1}$. The starting values of the symmetry-breaking LECs $b_0,b_D,b_F$ are chosen to be consistent with ground-state octet baryon masses and $\sigma_{\pi N}$ at the next-to-leading chiral order. In the fit, small variations around these values are allowed. We found that this stabilizes the fit considerably. At every step of the fit routine, we impose an analyticity constraint on the scattering amplitude, disregarding solutions with poles on the first Riemann sheet closer than 150~MeV to the real axis.

The obtained best fit has an order of magnitude smaller $\chi^2_{\rm dof}$ than the next best fits. Its parameters are collected in the left panel of Tab.~\ref{tab:fit-results}. The experimental data is well reproduced by the model as shown in Figs.~\ref{pic:TOT-CS} and \ref{pic:DIF-CS} in the Appendix. In Fig.~\ref{pic:TOT-CS} also the s-wave cross section is shown. The p-wave contribution to the total cross section is very small but p-wave as such still has a major influence on the fit through the differential cross section. The major contribution to the $\chi^2_{\rm dof}$ originates from the total cross sections data (a), see values in parenthesis. All model parameters are of natural size with $1\sigma$-uncertainties determined in a re-sampling procedure. The quantity $|\Sigma|$ shows the generalized variance that is proportional to the volume of the error ellipse of the 20-dimensional parameter space which can serve as a bulk measure of how well the data restrict the parameters. In particular, in the region with confidence $1-\alpha$ around the best fit parameters ($p$), the volume of ellipse reads $\frac{(2\pi)^{p/2}}{p\Gamma(\frac{p}{s})}(\chi^2_{p,\alpha})^{p/2}|\Sigma|^{1/2}$.
We will set the quoted number ($0.106$~GeV$^{-14}$) into relation when discussing the stability of results in the following. Furthermore, exploring the covariance matrix, we have not found any obvious separation between various parameter groups.

The interaction kernel~\eqref{eqn:potential} contains explicit (linear) dependence on the (cosine of) scattering angle, and so does the solution of the dynamical coupled-channel equation~\eqref{eqn:BSE}. The best fit solution is projected to the s- and p-waves, denoted as in Ref.~\cite{Hohler:1984ux} by $f^I_{L\pm}$, indicating total angular momentum $J=L\!\pm\!\nicefrac{1}{2}$ and isospin $I$. This is depicted in Fig.~\ref{fig:amplitudes} for the $\bar KN$ channel including $1\sigma$ error bands up to energies covered by the considered data base (see dots above Fig.~\ref{fig:amplitudes}). Around the $\bar KN$ threshold the dominant contribution is due to the s-wave amplitudes with a clear resonant behavior in the sub-threshold region in the isoscalar channel.

To identify the resonance parameters the amplitudes $f^I_{L\pm}$ are analytically continued to the complex energy-plane on the second Riemann sheet (II.RS). Indeed, on the II.RS connected to the physical one between the $\pi\Sigma$ and $\bar KN$ thresholds we find two poles in the $0(1/2^-)$ sector, associated with the $\Lambda(1405)$, and one in the $1(1/2^+)$ channel, referred in the following to as $\Sigma(1380)1/2^+$.

The resonance parameters are collected in the right panel of Tab.~\ref{tab:fit-results}, including the magnitudes of the couplings to the meson-baryon channels {${g^2:=|\lim_{W\to W*}(W-W^*)f^I_{L\pm}(W)|}$} for the pole position $W^*$. As expected, the narrow pole of the $\Lambda(1405)$ couples dominantly to $\bar KN$, while the broad one more to the $\pi\Sigma$ channels. The new $\Sigma(1380)1/2^+$ state couples mostly to the $\pi^0\Lambda$ channel.

\begin{figure*}[t!]
\includegraphics[width=0.49\linewidth, trim=0 0 0 0]{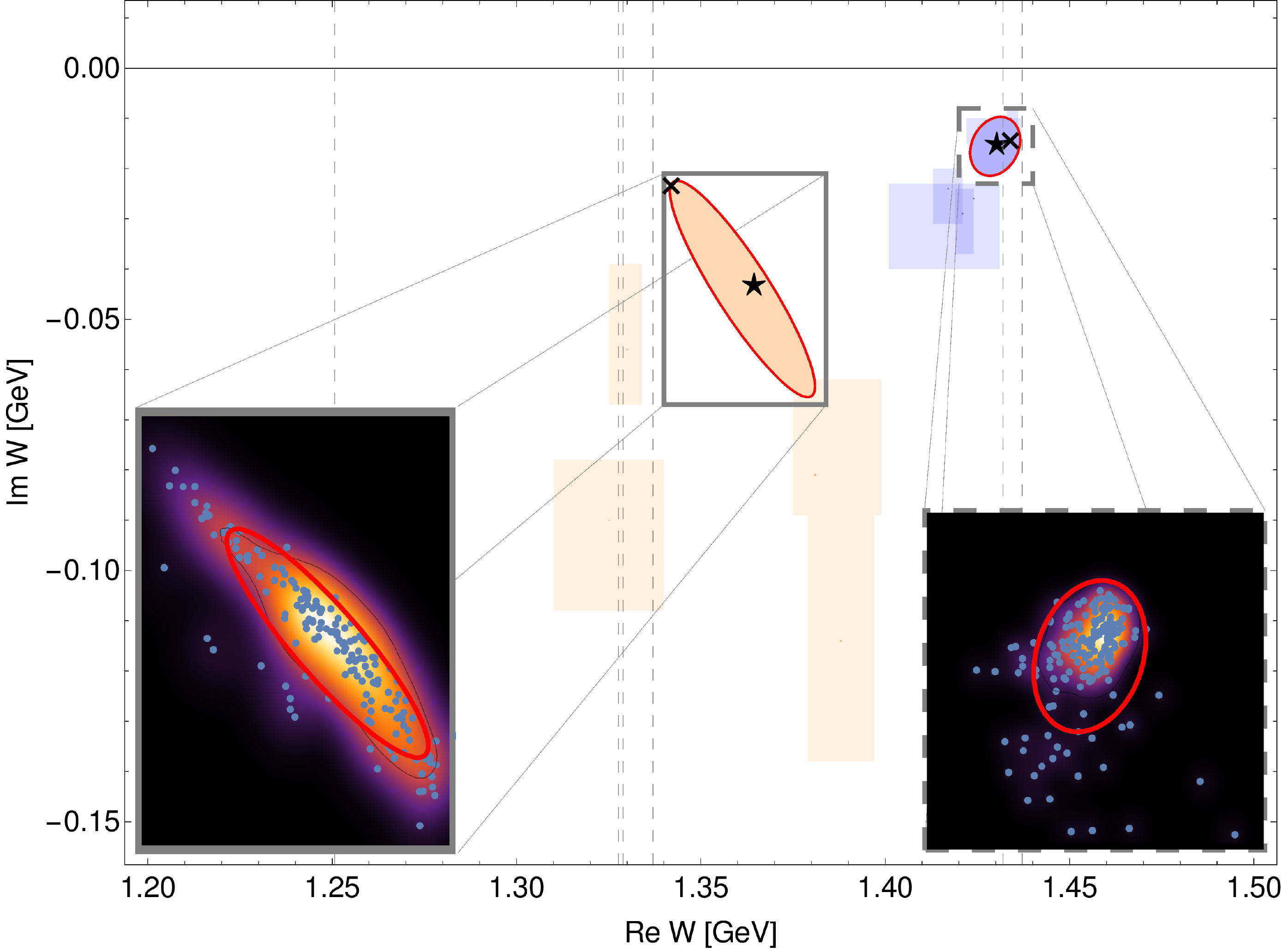}
\includegraphics[width=0.49\linewidth, trim=0 0 0 0]{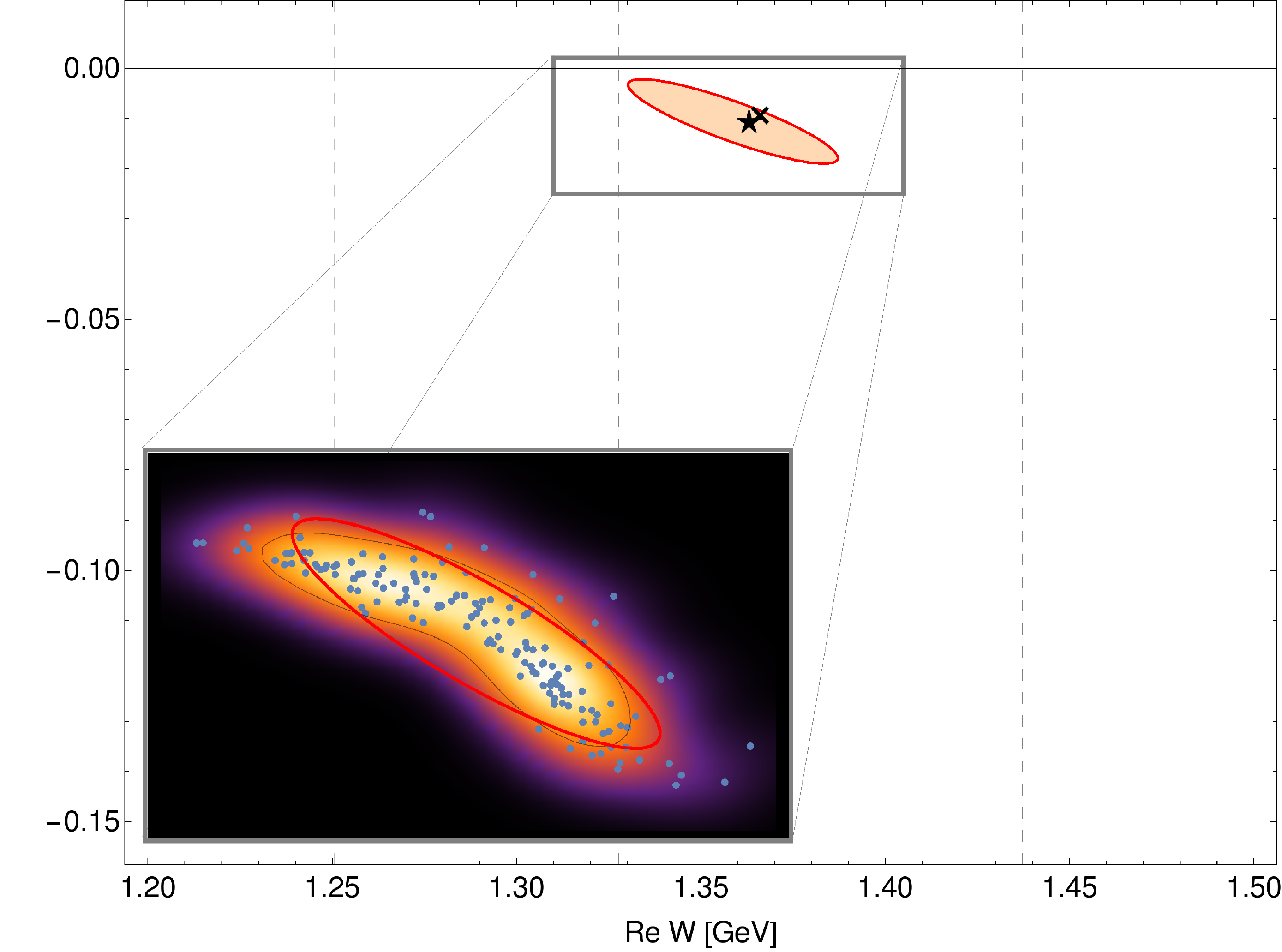}
\caption{
\label{fig:poles}
Pole positions (black stars at the centers of the ellipses) for the $I(J^P)=0(1/2^-)$ (left) and  $1(1/2^+)$ (right) channels. The error ellipses are from a re-sampling procedure shown explicitly in the corresponding insets. The shaded rectangles show the predictions of Refs.~\cite{Mai:2014xna,Ikeda:2011pi,Guo:2012vv} for the narrow (blue) and broad (orange) pole of $\Lambda(1405)$. On the left, the black crosses denote the locations of the $\Lambda(1405)$ poles when the $\Sigma(1385)3/2^+$ is included in the model (see ``Fifth'' in the main text). In that case, the $I(J^P)=1(1/2^+)$ pole (right, star) disappears and is replaced by the $\Sigma(1385)3/2^+$ (right, cross).
}
\end{figure*}

The resonances poles in the complex energy-plane are depicted and compared with the outcome of other models in Fig.~\ref{fig:poles}. Note that all these models are purely s-wave approaches, and do not include the differential cross section data (b) that bears information on the partial-wave content. This is in line with the observation that, while there is a good agreement on the parameters of both poles in the s-wave with these models, the new p-wave state has not been found in any previous analyses.

\paragraph{\textbf{Stability of results}}

The predictions for the both poles of the $\Lambda(1405)1/2^-$ are well in agreement with other modern approaches~\cite{Guo:2012vv,Ikeda:2011pi,Mai:2014xna}. 
The new p-wave $\Sigma(1380)1/2^+$-resonance has similar mass, width, and branching ratios as the decuplet $\Sigma(1385)3/2^+$ but different total angular momentum and is not listed in the PDG~\cite{Tanabashi:2018oca}. Before discussing the possible origin of this spurious state we perform several statistical and phenomenological tests to confirm its existence within the present model. 

\noindent
\underline{First:}
The $\pi\Sigma$-invariant mass distribution~\cite{Hemingway:1984pz} from the reaction ${K^-p\!\to\!\Sigma(1660)^+\pi^-\!\to\!((\pi^-\Sigma^+)\pi^+)\pi^-}$ can be used for an \textit{a-posteriori} test of the obtained solution. For this and following Ref.~\cite{Oller:2000fj} we assume energy-independent production and decay vertices of the $\Sigma(1660)1/2^+$. Fitting the associated constants, while not altering the meson-baryon scattering amplitude in the final state, we obtain a good fit with $\chi^2_{\rm pp}=0.89$ (per data point). The result is shown in Fig.~\ref{pic:HEMINGWAY} in the Appendix.

\noindent
\underline{Second:}
A similar test of our scattering amplitude can be performed using the $\pi\Sigma$-invariant mass distributions~\cite{Moriya:2013eb} of the reactions $\gamma p\to K^+(\pi\Sigma)$ recently measured with CLAS at Jefferson Lab. Following Refs.~\cite{Roca:2013av,Mai:2014xna} we obtain a good fit of the data with $\chi^2_{\rm pp}=1.07$ (see Fig.~\ref{pic:CLAS} in the Appendix), fitting again only the generic couplings $\gamma N\to K^+\mathcal{S}_i$.

\noindent
\underline{Third:}
The differential cross section data (b) contributes with 30\% to the full $\chi^2_{\rm dof}$, see Tab.~\ref{tab:fit-results} that also contains the total cross section data (a) and threshold decay ratios (c). Additionally, we test the impact of the these data by excluding it. Indeed, the generalized variance $|\Sigma|$, that provides a bulk number of how well data restrict parameters, increases from $|\Sigma_{a,b,c}|=0.106$~GeV$^{-14}$ to $|\Sigma_{a,c}|=9.74$~GeV$^{-14}$. In other words, the parameters of the model are significantly less well determined when omitting the differential cross section data, demonstrating their relevance. Furthermore, the $1(1/2^+)$ pole in the complex energy plane disappears -- the existence of the $\Sigma(1380)1/2^+$ seems indeed tied to the differential cross section data.

\noindent
\underline{Fourth:}
To investigate the stability of the new p-wave resonance further, we use the least absolute shrinkage and selection operator  (LASSO)~\cite{hastie_hastie_tibshirani_friedman_2001} to remove the $\Sigma(1380)1/2^+$. In this we add a penalty term $\lambda\,\int_{1.29\,\text{GeV}}^{1.46\,\text{GeV}}|\partial^2/\partial W^2 \,f^{I=1}_{1-}(W))| dW$  to the total $\chi^2$. Minimizing this function at $\lambda=0$ and including data (a,b,c) we obtain back our previous result. Subsequently increasing the value of $\lambda$ and re-fitting, the curvature of the amplitude between the $\pi\Sigma$ and $\bar KN$ thresholds decreases forcing the p-wave resonance to disappear into the complex plane as shown in Fig.~\ref{fig:LASSO}. The figure also reveals that the description of the threshold data (c) deteriorates most visibly as the pole disappears. Of course, this is not \emph{directly} due to the disappearance of the pole, but related to it through parameters $b_i$ and $a_i$.

The quantity $\lambda_{\rm crit}$ is the minimum $\lambda$, at which $|\partial^2/\partial W^2 \,f^{I=1}_{1-}(W)|$ is close to zero for ${W\approx1380}$~MeV. The difference in quality of the fit of the data when $\lambda=0$ (our best fit) with the fit of the data at $\lambda_{\rm crit}$ measures the stability of the resonance. The likelihood ratio of the fit at $\lambda=\lambda_{\rm crit}$ to the fit at $\lambda=0$ is $4.3\times 10^{-28}$ which shows that the data description indeed deteriorates significantly in the absence of the resonance. In particular, the prediction of kaonic hydrogen data~\cite{Bazzi:2011zj} is more than 2 standard deviations away from the experimental value at $\lambda_{\rm crit}$.

\noindent
\underline{Fifth:}
The p-wave decuplet $\Sigma(1385)$ resonance in the $J^P=3/2^+$ partial wave is absent in our model because the NLO terms do not fully describe the $J^P=3/2^+$ sector.  On the other hand, the slope of the differential cross section indicates the size of the p-wave contribution (Fig.~\ref{pic:DIF-CS}) but is blind to the total angular momentum. It is then obvious to ask if the newly found $\Sigma(1380)1/2^+$ is only generated to account for the absence of the $\Sigma(1385)3/2^+$.
This could indeed be the case because the $\Sigma(1380)1/2^+$ exhibits  very similar pole position and branching ratios as the $\Sigma(1385)3/2^+$.

\begin{table}[htp]
\centering

\resizebox{\columnwidth}{!}{\begin{tabular}{| c | c | c | c | c | c | c | c | c | c | c | c | c | c }
\hline
  $b_1$ & $b_2$ & $b_3$ & $b_4$ & $b_5$ & $b_6$ & $b_7$ & $b_8$ & $b_9$ & $b_{10}$ & $b_{11}$  \\
-0.34 & -0.06 & -0.59 & 0.15 & 0.14 & -0.61 &
-0.97 & 0.04 & 0.37 & 0.16 & 0.32 \\
\hline
$b_0$ & $b_D$ & $b_F$ & $a_1$ & $a_2$ & $a_3$ & $a_4$ & $a_5$ & $a_6$ & $g$ & $m_0$\\
-0.46 &
0.13 & -0.27 & 0.32 & 5.48 & 0.27 & -1.07 & -0.96 & -1.44 & 1.50 & 1.52\\
\hline
\end{tabular}}
\bigskip
\resizebox{\columnwidth}{!}{\begin{tabular}{| c  c  c  c  c | }

$I(J^P)$& &$0(1/2^-)$&$0(1/2^-)$&$1(1/2^+)$\\
Pole Positions [GeV] & & 1.3644 -0.0431i & 1.4302 -0.0151i & 1.366-0.009i \\
\hline
\end{tabular}}
\caption{Best fit parameters in GeV and pole positions found when an explicit $\Sigma(1385)3/2^+ $ state is included in the fit. $g$ and $m_0$ are the new parameters for this test.} 
\label{tab:newtable}
\end{table}

To test this hypothesis, we explicitly include it in order to replace the new  $\Sigma(1380)1/2^+$. The two-potential formalism allows the addition of resonances to an amplitude without spoiling unitarity~\cite{Ronchen:2012eg}.
This is achieved by adding the term
$f_{1+}^P=\Gamma^*[W-m_0-\gamma^T I \Gamma^*]^{-1}\Gamma^T$ to $f_{1+}$ where T indicates the transpose of the ``dressed vertex'' $\Gamma= \gamma+  \gamma I f_{1+}$ which is a vector in channel space. The ``bare vertex'' vector $\gamma$ has components $\gamma_i=g\,g_i$ with one free fit parameter $g$ and the relative decay strengths $g_i$ to channels $\pi\Lambda$, $\pi\Sigma$, and $\bar{K}N$ fixed by the Lagrangian of Ref.~\cite{Butler:1992pn}. Furthermore, $I$ is the diagonal matrix of meson-baryon loop functions from the integration in Eq.~(\ref{eqn:BSE}). The bare mass $m_0$ and bare coupling $g$ are new fit parameters.

 We then fit the previous free parameters as well as the two new ones shown in Tab.~\ref{tab:newtable}. These parameters give  a  $\Sigma(1385)3/2^+$ pole position of $W=1.366-0.009i$~GeV. This value is 16~MeV below the Particle Data Group average value of 1382.8~MeV, but it is within the error ellipse of the spurious structure we observe (see the cross in Fig.~\ref{fig:poles}). The changed pole positions of the $\Lambda(1405)$, also indicated by crosses, lie within the error ellipses of the previous solution.  The ratio of partial decay widths for the $\Sigma(1385)3/2^+$ is predicted as $\Gamma(\Sigma\pi)/\Gamma(\Lambda\pi)=0.183$ compared to the PDG value of $\Gamma(\Sigma\pi)/\Gamma(\Lambda\pi)=0.135$~\cite{Tanabashi:2018oca}.  This test causes the p-wave $\Sigma(1380)1/2^+$ observed in our previous fit to disappear. 
 
 Therefore, the results of this test indicate that the $\Sigma(1380)1/2^+$ is dynamically generated by the free parameters just to account for the missing $\Sigma(1385)3/2^+$. Indeed, no $\Sigma(1380)1/2^+$ is found in Lattice QCD calculations~\cite{Edwards:2012fx, Engel:2013ig, Engel:2010my}.
 


It should be noted that a more rigorous analysis would require the inclusion of the $\Sigma(1385)3/2^+$ as a spin$-\nicefrac{3}{2}$ field not only in the $s$ but also the $u$-channel in  Eq.~\eqref{eqn:potential}. This is beyond the scope of this work. Similar considerations apply to the ground-state $\Lambda$ and $\Sigma$ in the $s$- and $u$-channels.


\paragraph{\textbf{Conclusion}}
For the first time the available low-energy differential cross section data have been included in a chiral unitary coupled-channel model with fully dynamically generated s- and p-waves, i.e., all waves being parametrized only by low-energy and regularization constants. Parameters of both $\Lambda(1405)1/2^-$ poles are determined in this approach with improved stability. Additionally, and confirmed by multiple (statistical and phenomenological) tests, the data require the existence of a p-wave pole but are otherwise insensitive to the total angular momentum. Thus, the pole is predicted in the only available p-wave with $J^P=1/2^+$ but a number of tests show that this dynamically generated pole accounts for the absence of the $\Sigma(1385)3/2^+$ in the model.

\paragraph{Acknowledgments}
MM thanks the German Research Association (MA 7156/1) for the financial support. MD acknowledges support by the National Science Foundation (grant nos. 1415459 and PHY-1452055) and by the U.S. Department of Energy, Office of Science, Office of Nuclear Physics under contract no. DE-AC05-06OR23177 and grant no. DE-SC0016582.

\bibliography{MM-ref.bib}

\begin{figure*}[h!]
\includegraphics[width=0.9\linewidth, trim = 0 2cm 0 2cm ]{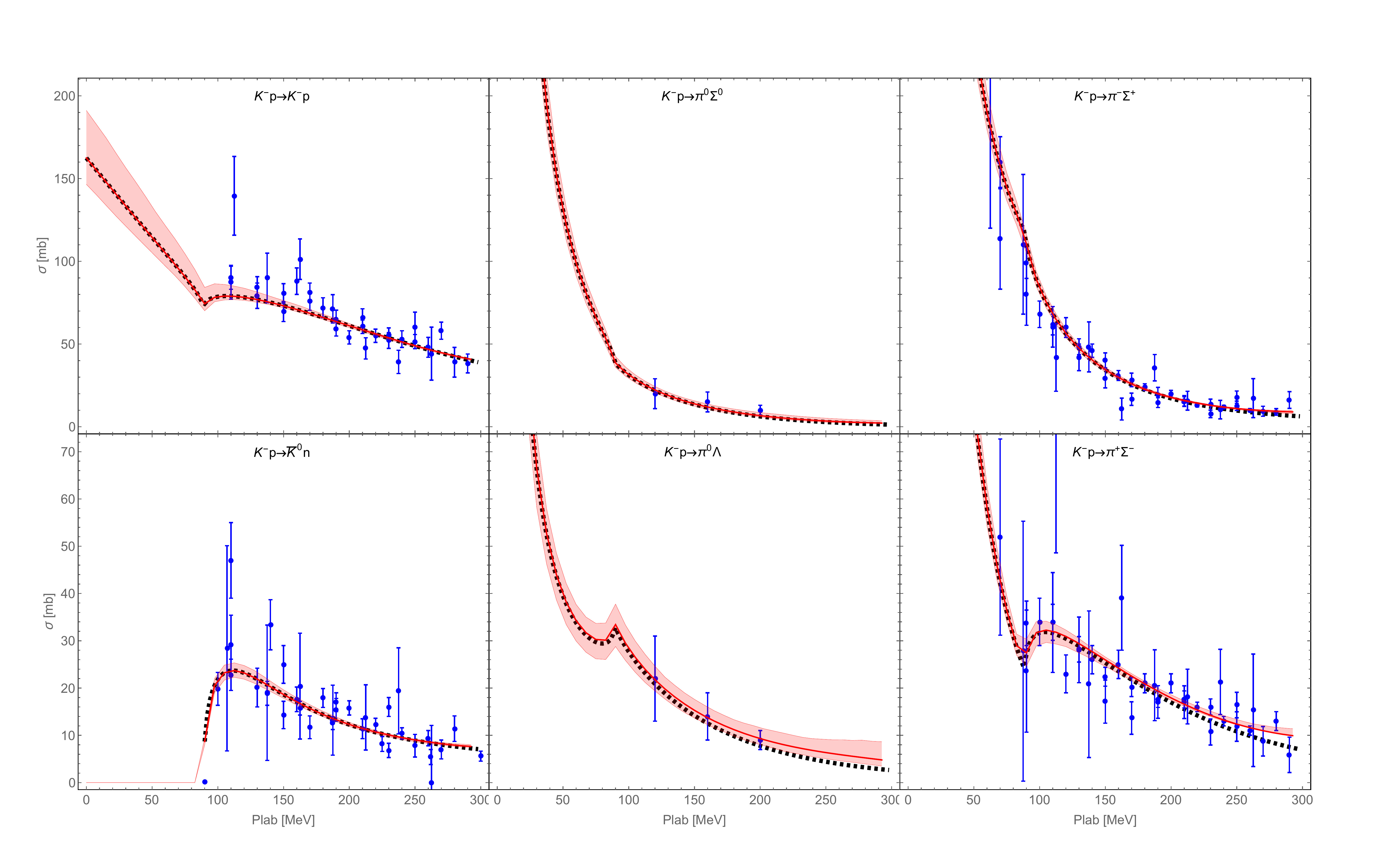}
\caption{Total cross sections~\cite{Nowak:1978au,Watson:1963zz,Ciborowski:1982et,Humphrey:1962zz} fitted by the model described in the main part of the manuscript. Error bands represent the $1\sigma$ uncertainty determined in a re-sampling procedure. The dashed black line shows the contribution of the s-wave part of the amplitude only.}
\label{pic:TOT-CS}
\end{figure*}
\begin{figure*}[h!]
\includegraphics[width=0.9\linewidth, trim = 0 2cm 0 2cm ]{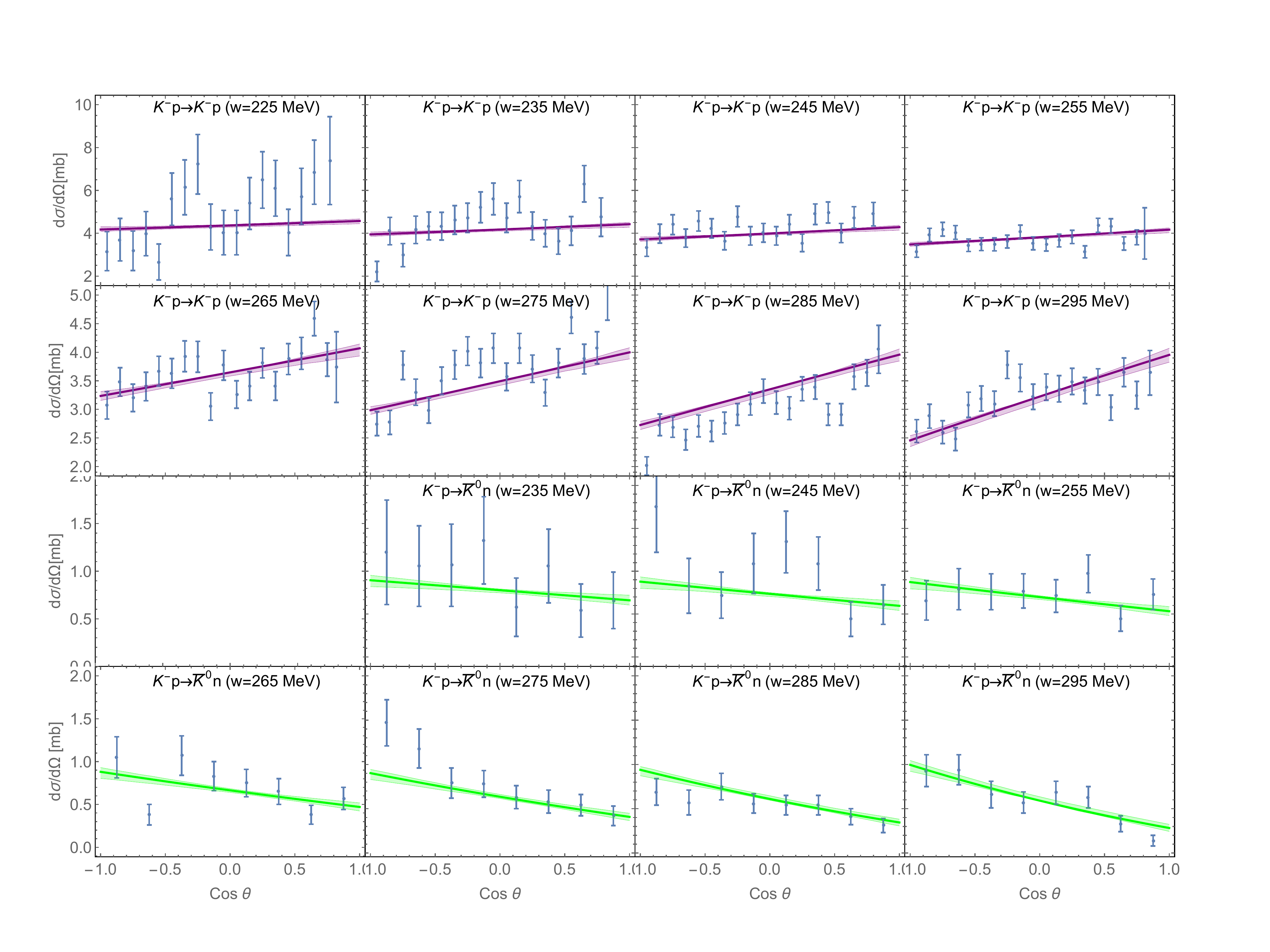}
\caption{Differential cross sections~\cite{Mast:1975pv} fitted by the model described in the main part of the manuscript. Error bands represent $1\sigma$ uncertainty determined in a re-sampling procedure.}
\label{pic:DIF-CS}
\end{figure*}

\begin{figure*}
\includegraphics[width=0.88\linewidth]{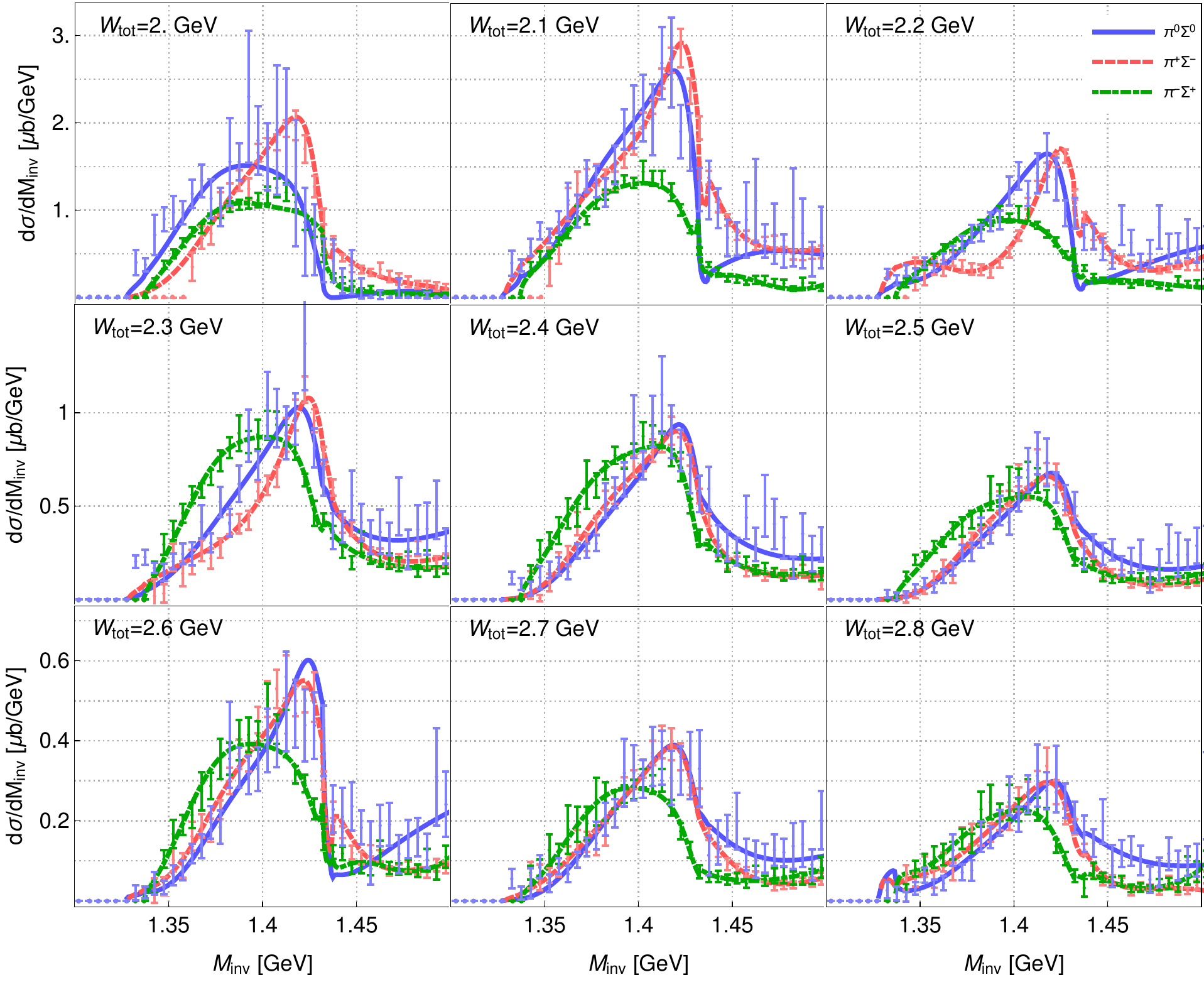}
\caption{
Fit ($\chi^2_{\rm pp}=1.07$) to the $\pi\Sigma$ invariant mass distribution ($M_{\rm inv}$) from $\gamma p\to K^+ (\pi\Sigma)$ reaction~\cite{Moriya:2013eb}. The model for the reaction is taken from Ref.~\cite{Mai:2014xna}, where only generic couplings $\gamma p\to K^+ \mathcal{S}$ are fitted to the data at each measured total energy $W_{\rm tot}$. The meson-baryon scattering amplitude in the final state is taken from the fit to the scattering data. Only the best fits are shown.
}
\label{pic:CLAS}
\end{figure*}

\begin{figure*}
\nocaption
\begin{subfigure}{0.49\linewidth}
\includegraphics[width=0.99\linewidth]{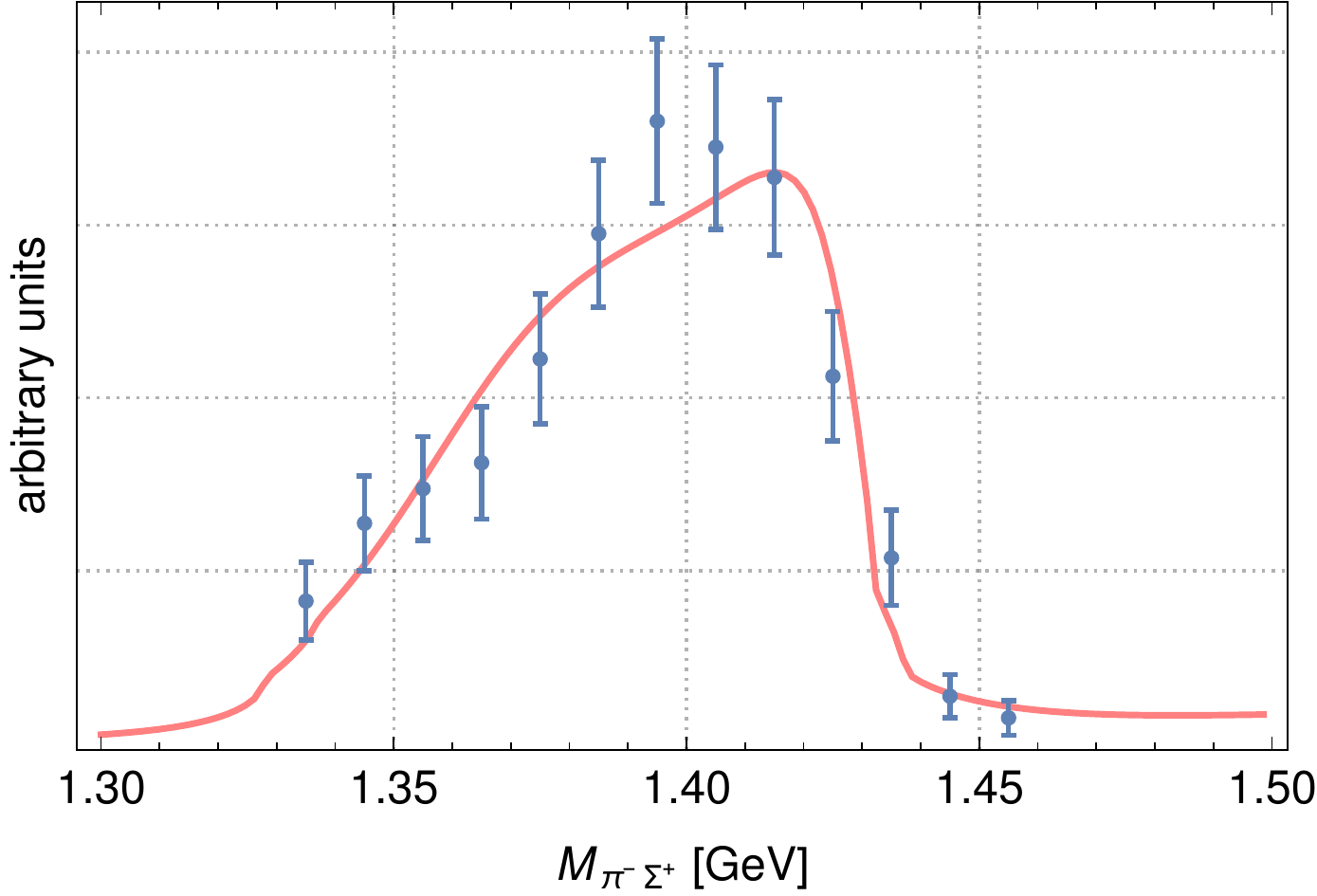}
\caption{Fit of the generic couplings $K^-p\to\Sigma(1660)\pi^-$ and  $\Sigma(1660)\to(\pi^-\Sigma^+)\pi^+$ to the invariant mass distribution~\cite{Hemingway:1984pz} in arbitrary units. The final state interaction is taken from the best fit to the scattering data as described in the main part of the manuscript. Only the best fit is shown.
}\label{pic:HEMINGWAY}
\end{subfigure}
%
\begin{subfigure}{0.49\linewidth}
\includegraphics[width=0.99\linewidth]{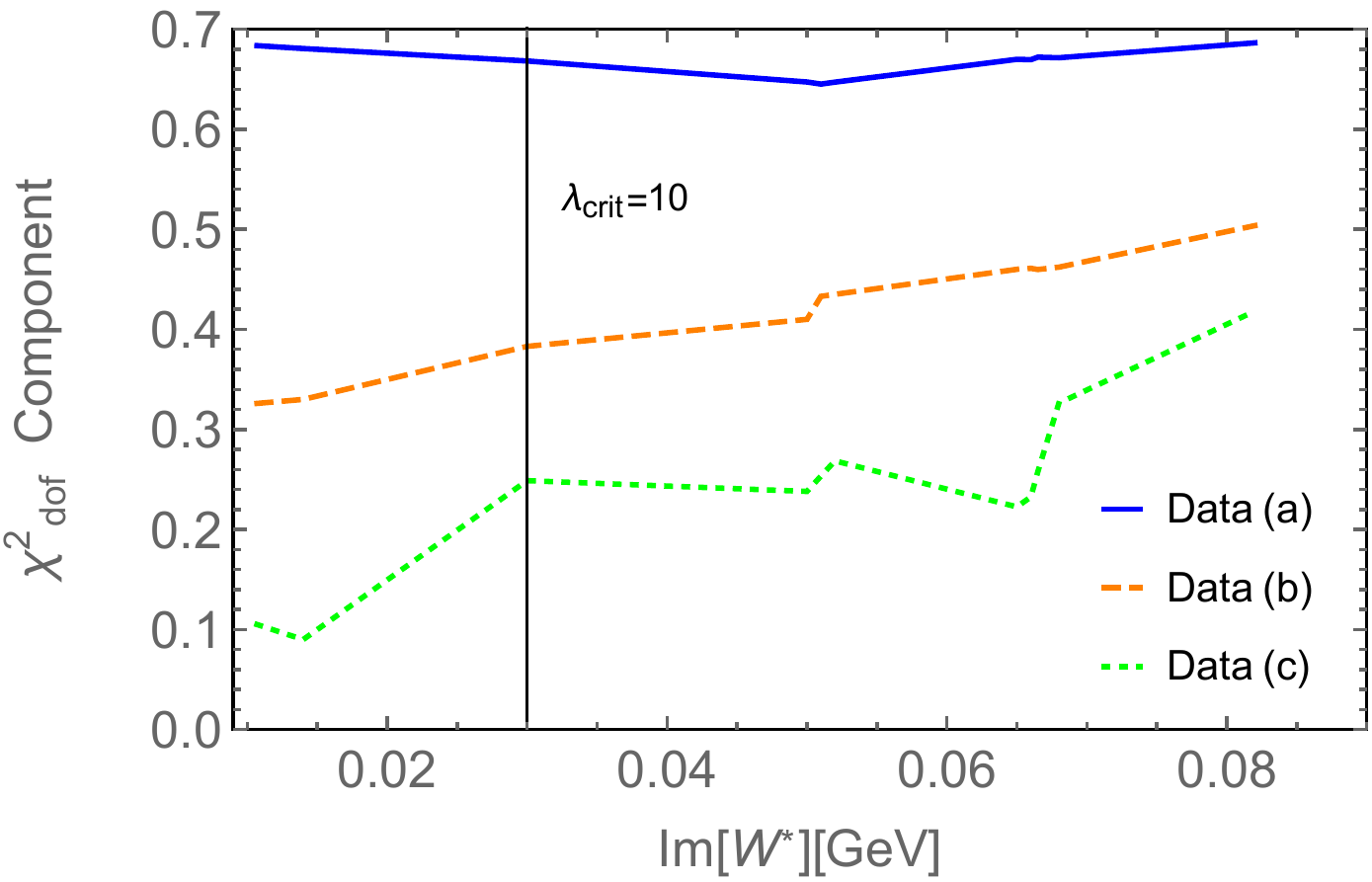}
\caption{
Dependence of the components of the $\chi^2_{\rm dof}$ from each data set on the distance of the $W^*$ pole from the real axis, parametrized with the penalty $\lambda$. The sum of all three components equals $\chi^2_{\rm dof}$ which increases with the imaginary position of the pole at $W=W^*$.
The contribution to the $\chi^2$ from the penalty is subtracted in all cases.
}\label{fig:LASSO}
\end{subfigure}
\end{figure*}

\end{document}